\documentclass[11pt]{article}
\usepackage{a4wide}
\usepackage{amsmath, amssymb, amscd, amsthm, amsfonts}
\usepackage[utf8]{inputenc}
\usepackage{graphicx}
\usepackage{hyperref}
\usepackage{cite}
\usepackage[dvipsnames]{xcolor}

\usepackage{lmodern}
\usepackage{pgfplots}
\pgfplotsset{compat=1.8}

\newcommand{\ket}[1]{\left| #1 \right\rangle} %ket%
\newcommand{\bra}[1]{\left\langle #1 \right|} %bra%

\newcommand{\be}{\begin{equation}}
\newcommand{\ee}{\end{equation}}
\newcommand{\ba}{\begin{align}}
\newcommand{\ea}{\end{align}}

\begin{document}
{\renewcommand{\thefootnote}{\fnsymbol{footnote}}
		
\begin{center}
{\LARGE TCC bounds on the static patch of de Sitter space} 
\vspace{1.5em}
\\
Mattéo Blamart\footnote{e-mail address: {\tt matteo.blamart@mail.mcgill.ca}}, 
Samuel Laliberte\footnote{e-mail address: {\tt samuel.laliberte@mail.mcgill.ca}} and 
Robert Brandenberger\footnote{e-mail address: {\tt rhb@physics.mcgill.ca}}
\\
\vspace{1.5em}
Department of Physics, McGill University, Montr\'{e}al, QC, H3A 2T8, Canada

\vspace{1.5em}
\end{center}
}
	
\setcounter{footnote}{0}

\begin{abstract}
	\noindent Recently, Pei-Ming Ho and Hikaru Kawai \cite{HK} have argued that treating particles as wave packets can lead to a shutdown of Hawking radiation after a scrambling time in the case of Schwarzschild black holes.  This shutdown arises from viewing quantum field theory near the black hole horizon as an effective field theory, and imposing an appropriate UV cutoff.  We show that this effect is also present in the static patch of de Sitter space, leading to a shutdown of Gibbons-Hawking radiation at late times.  Assuming this effect is due to the breakdown of effective field theory, we obtain a bound $t \lesssim H^{-1} \ln (H^{-1} M_P)$ on the time scale of validity of effective field theory in de Sitter space, which matches with the predictions of the Trans-Planckian Censorship Conjecture.
\end{abstract}

\tableofcontents

\section{Introduction}

It is well known that, calculated an effective field theory (EFT) in the semiclassical approximation, black holes radiate thermally \cite{Hawking}.  Specifically, if matter is treated as a scalar field, then an observer at late times and far from the horizon will measure a thermal flux of scalar field quanta if the initial state is prepared as an appropriate vacuum state. In a similar way, observers at a point $x$ in de Sitter space will measure a thermal flux of scalar field quanta if the initial state is set up as a local vacuum state a Hubble horizon distance away from $x$ \cite{GH}. It is these fluctuations which are postulated to be the origin of density perturbations \cite{Mukh} and gravitational waves \cite{Starob} in an inflationary universe scenario \cite{Guth}.

In the context of inflationary cosmology, it was argued in \cite{Jerome} that there is a ``trans-Planckian problem'' for cosmological fluctuations if the period of inflation is long since in that case the physical wavelength of the fluctuations observed today was smaller than the Planck length at the beginning of inflation and thus in a region where the effective field theory of fluctuations becomes questionable.  Based on these considerations,  in \cite{Bedroya1} a ``Trans-Planckian Censorship Conjecture'' (TCC) was put forward according to which in no effective field theory emanating from superstring theory modes which initially were trans-Planckian could ever have exited the Hubble horizon, i.e.
\be
\frac{a(t_R)}{a(t_i)} l_{pl} \, < \, H^{-1}(t_R) \, ,
\ee
for any initial time $t_i$ and final time $t_R$. In the above, $a(t)$ is the cosmological scale factor, and $H(t)$ is the Hubble expansion rate at time $t$ (whose inverse is the Hubble horizon). As discussed in \cite{Bedroya2},  the TCC severely constrains viable inflationary models. In particular, canonical single scalar field inflation models with an energy scale $\eta > 10^{9} {\rm{GeV}}$ are ruled out. As argued in \cite{RBrev}, the TCC can be argued for independently of superstring theory, making use of unitarity arguments and consistency with the second law of thermodynamics (see \cite{Omar}).  

On the other hand,  there have been arguments to the effect that there is no trans-Planckian problem for inflationary cosmology (see e.g. \cite{Starob3, Dvali3, Burgess}). Based on the Einstein equivalence principle it is argued that on microscopic scales (in the context of cosmology this means on length scales much smaller than the Hubble radius) physics must reduce to that in Minkowski space-time.  From the point of view of quantum gravity, however, this argument is problematic.  Nevertheless, it would be useful to have more independent support for the TCC.

In the case of black holes, one can argue that a similar trans-Planckian problem arises (see e.g. \cite{Unruh, Jacobson} for some early work).  From the point of view of a semiclassical EFT analysis, the Hawking radiation \cite{Hawking} which an observer far outside the black hole horizon observes at a frequency $\omega$ (e.g. the frequency corresponding to the Hawking temperature) observes is associated with a blueshifted frequency $\Omega$ which becomes larger than the Planck scale sufficiently close to the black hole horizon. Hence, one may argue that the semi-classical analysis must break down. 
 
 Once again, it can be argued that there is no problem for the semiclassical description of Hawking radiation since there is a ``nice slice'',  a spatial hypersurface from whose point of view there is no divergence of the frequency close to the horizon (see e.g. \cite{nice}).  But this does not resolve the famous black hole information loss problem (see e.g. \cite{info} for early discussions of this problem) associated with Hawking radiation.  It has been argued \cite{Page} that the semiclassical analysis must break down at the ``Page time'', the time when the entropy of thermal Hawking radiation reaches half of the initial entropy of the black hole (as determined by the area of the initial black hole horizon).\footnote{Lately, significant progress has been made towards understanding black hole evolution after a Page time \cite{BHentropy}.  Similar analyses have also been carried out in de Sitter space, with comparable results \cite{dSislands}.}
 %{\color{red} I'm not sure if this paragraph still makes sense if we replace Page time by scrambling time.  We should think about this.}

In recent work, Pei-Ming Ho and Hikaru Kawai \cite{HK} considered the propagation of wave packets in a black hole background rather than considering only plane waves.  Working in the context of effective field theory and imposing a UV cutoff,  they find that Hawking radiation shuts off at the scrambling time.  More generally, they find that the spectrum of Hawking radiation after the scrambling time depends on physics which cannot be described in terms of the EFT.

We apply a similar analysis to the static patch dS,  and find that Gibbons-Hawking radiation shuts off after the TCC time scale (if there is an ultraviolet cutoff), and more generally cannot be described by the EFT beyond that time.  Besides the intrinsic interest of this result,  it points to an intriguing analogy between the black hole scrambling time and the cosmological TCC time scale.

%In de Sitter space, an observer moving along a time-like geodesic observes a thermal bath of particles with a characteristic temperature $T = (2 \pi l)^{-1}$, where $l$ is the de Sitter length scale.  %This effect is similar to Hawking radiation in the sense that 

This document is structured as follows.  In section \ref{sec:rev}, we review Gibbons-Hawking radiation in de Sitter space by using a conventional approach, where particles on a curved background behave like plane waves.  In section \ref{sec:wp}, we define semi-classical wave packets and describe how they can arise as fluctuations of quantum field fields.  Finally, in section \ref{sec:shut}, we show how Gibbons-Hawking radiation coming from particles that behave as semi-classical wave packets becomes sensitive to UV cutoffs after a scrambling time $t_{sc} = H^{-1} \ln ( H^{-1} M_p )$, leading to a breakdown of effective field theory.

\section{Review of Gibbons-Hawking radiation in de Sitter space}
\label{sec:rev}

To see how thermal radiation can be shut down after a finite amount of time, let us start by reviewing the derivation of Gibbons-Hawking radiation when particles are treated as plane waves in a curved background.  We will roughly follow the approach presented in \cite{Brout:1995rd} and work in the static patch of de Sitter space where the metric takes the form
\be
ds^2 = - f(r) dt^2 + \frac{dr^2}{f(r)} + r^2 d\Omega_2^2 \quad , \quad f(r) = 1 - r^2/l^2 \, ,
\ee
where $l$ is the radius of the de Sitter horizon.  This metric can be rewritten in the more convenient Eddington-Finkelstein frame (see \cite{Spradlin:2001pw} for a review) where the metric takes the form
\be
ds^2 = - f(u,v) dudv + r(u,v)^2 d\Omega^2_2 \, ,
\ee
by defining a radial coordinates $r^* = \text{Arctanh} \left( r/l \right)$ and light cone-coordinates $u = t - r_*$ and $v = t + r_*$.  %In this frame, the future and past horizons respectively coincide with the $v \rightarrow \infty$ and $u \rightarrow - \infty$ limit.  
Another useful frame is the Kruskal coordinate system.  One can move from the Eddington-Finkelstein frame to the Kruskal frame using the change of coordinates
\be
U(u) = l e^{u/l} \quad , \quad V(v) = - l e^{-v/l} \, ,
\ee
in which case the space-time metric then takes the form
\be
ds^2 \approx - 4 dU dV  + r^2(U,V) d\Omega^2_2 \, ,
\ee
in the past and future horizon limits $u \rightarrow -\infty$ , $v \rightarrow \infty$.  %Here, the future and past horizon correspond respectively to the $V \rightarrow 0$ and $U \rightarrow 0$ limits.  
A useful feature of the Kruskal coordinate system is that space-time looks flat in the near horizon limit.  In comparison, the Eddington-Finkelstein coordinates look like those of an accelerated observer in the "flat" Kruskal background.  Consequently, one can derive Gibbons-Hawking radiation in de Sitter space by computing the spectrum of particles observed by an Eddington-Finkelstein observer in the Kruskal vacuum in the past and future horizon limits.  %Here, will be concerned about radiation near the future and past horizons.  Near the future horizon, Gibbons-Hawking radiation will be computed from left-moving moves

To do this, let us consider s-wave modes of a massless scalar field $\phi$ in the static patch of de Sitter space.  The mode decomposition of $\phi$ for s-wave modes in Kruskal and Eddington-Finkelstein coordinates can be written as \footnote{ This expansion can be modified to include angular momentum modes (see \cite{Bazarov:2021rrb} for a more detailed analysis).  For the present analysis, we will only be interested in s-waves for which the present mode expansion applies.}
\begin{align}
	\phi & = \int_0^{\infty} \frac{d\Omega}{\sqrt{4\pi\Omega} } \left( a_\Omega e^{-i \Omega U} + a^\dagger_\Omega e^{i \Omega U} + \tilde{a}_\Omega e^{-i \Omega V} + \tilde{a}^\dagger_\Omega e^{i \Omega V}\right) \\
	& = \int_0^{\infty} \frac{d\omega}{ \sqrt{4\pi\omega} } \left( b_\omega e^{-i \omega u} + b^\dagger_\omega e^{i \omega u} + \tilde{b}_\omega e^{-i \omega v} + \tilde{b}^\dagger_\omega e^{i \omega v}\right) \, . 
\end{align}
Here, $a_\Omega , \tilde{a}_\Omega$ and $b_{\omega} , \tilde{b}_\omega$ are annihilation operators defined by
\be
a_{\Omega} |0\rangle_a = \tilde{a}_{\Omega} |0\rangle_a = 0 \quad , \quad b_{\omega} |0\rangle_b = \tilde{b}_{\omega} |0\rangle_b = 0
\ee
where $|0 \rangle_a$ is the Kruskal coordinates vacuum state, and $|0\rangle_b$ is the Eddington-Finkelstein coordinates vacuum state.  Similar creation operators can be obtained by taking the hermitian conjugate of the annihilation operators.  The creation-annihilation of the two frames can be related by the Bogoliubov transformations
\begin{align}
	b_{\omega} & = \int_0^{\infty} d\Omega \left( \alpha_{\omega \Omega} a_{\Omega} + \beta_{\omega \Omega} a^\dagger_\Omega \right) \quad & \quad \tilde{b}_{\omega} & = \int_0^{\infty} d\Omega \left( \tilde{\alpha}_{\omega \Omega} \tilde{a}_{\Omega} + \tilde{\beta}_{\omega \Omega} \tilde{a}^\dagger_\Omega \right) \\
	b^\dagger_{\omega} & = \int_0^{\infty} d\Omega \left( \alpha^*_{\omega \Omega} a^\dagger_{\Omega} + \beta^*_{\omega \Omega} a_\Omega \right) \quad & \quad  \tilde{b}^\dagger_{\omega} & = \int_0^{\infty} d\Omega \left( \tilde{\alpha}^*_{\omega \Omega} \tilde{a}^\dagger_{\Omega} + \tilde{\beta}^*_{\omega \Omega} \tilde{a}_\Omega \right) \, ,
\end{align}
for appropriate Bogoliubov coefficients $\alpha_{\omega \Omega},\beta_{\omega \Omega}$ and $\tilde{\alpha}_{\omega \Omega},\tilde{\beta}_{\omega \Omega}$.  %An advantage of working in light-cone coordinates is that left moving modes (those depending on $v$ and $V$) and the right moving modes (those depending on $u$ and $U$) do not mix under Bogolibuv transformation.

\begin{figure}[h]
	\centering
        \includegraphics[width=6cm]{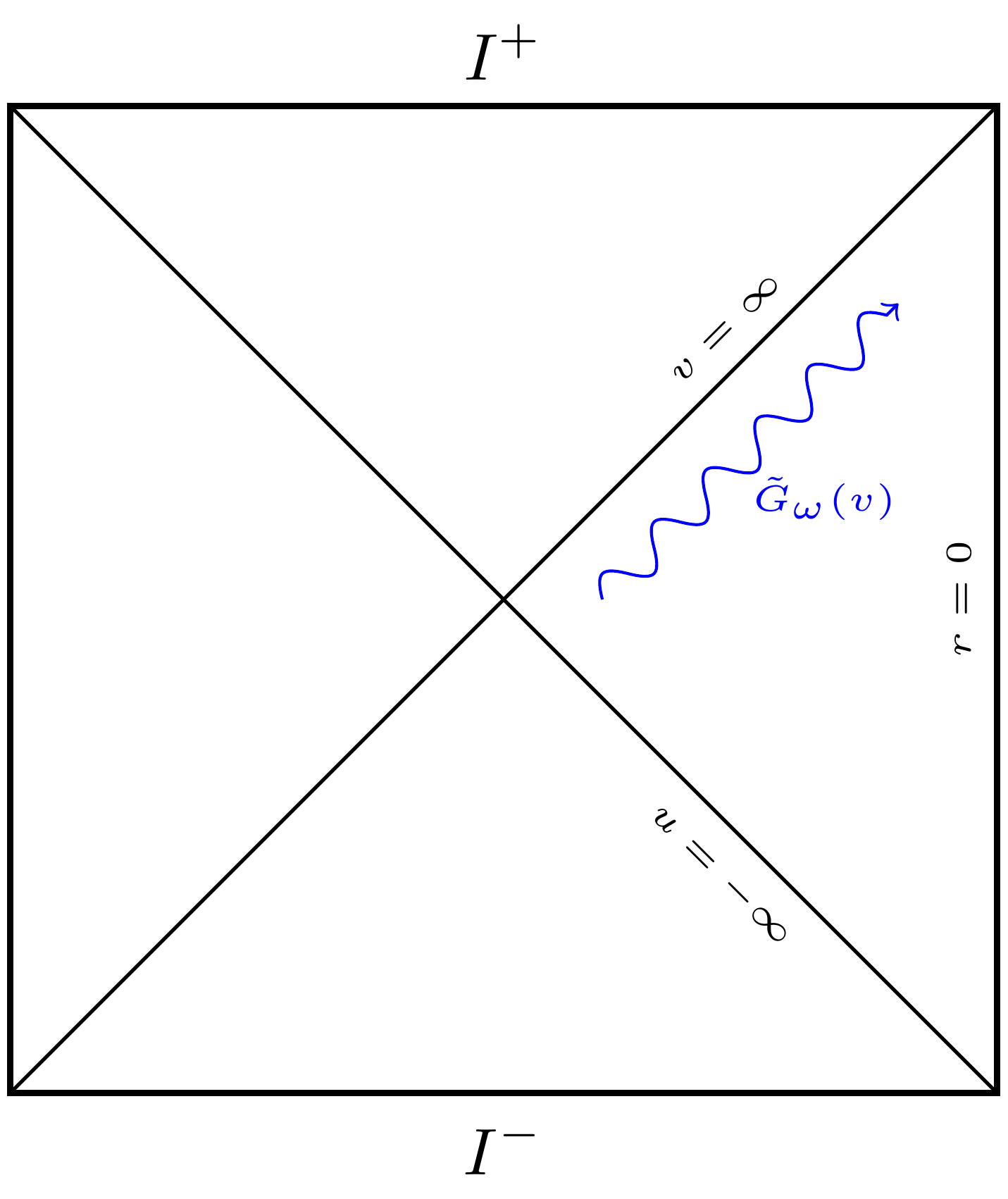} \hspace{1cm}
        \includegraphics[width=6cm]{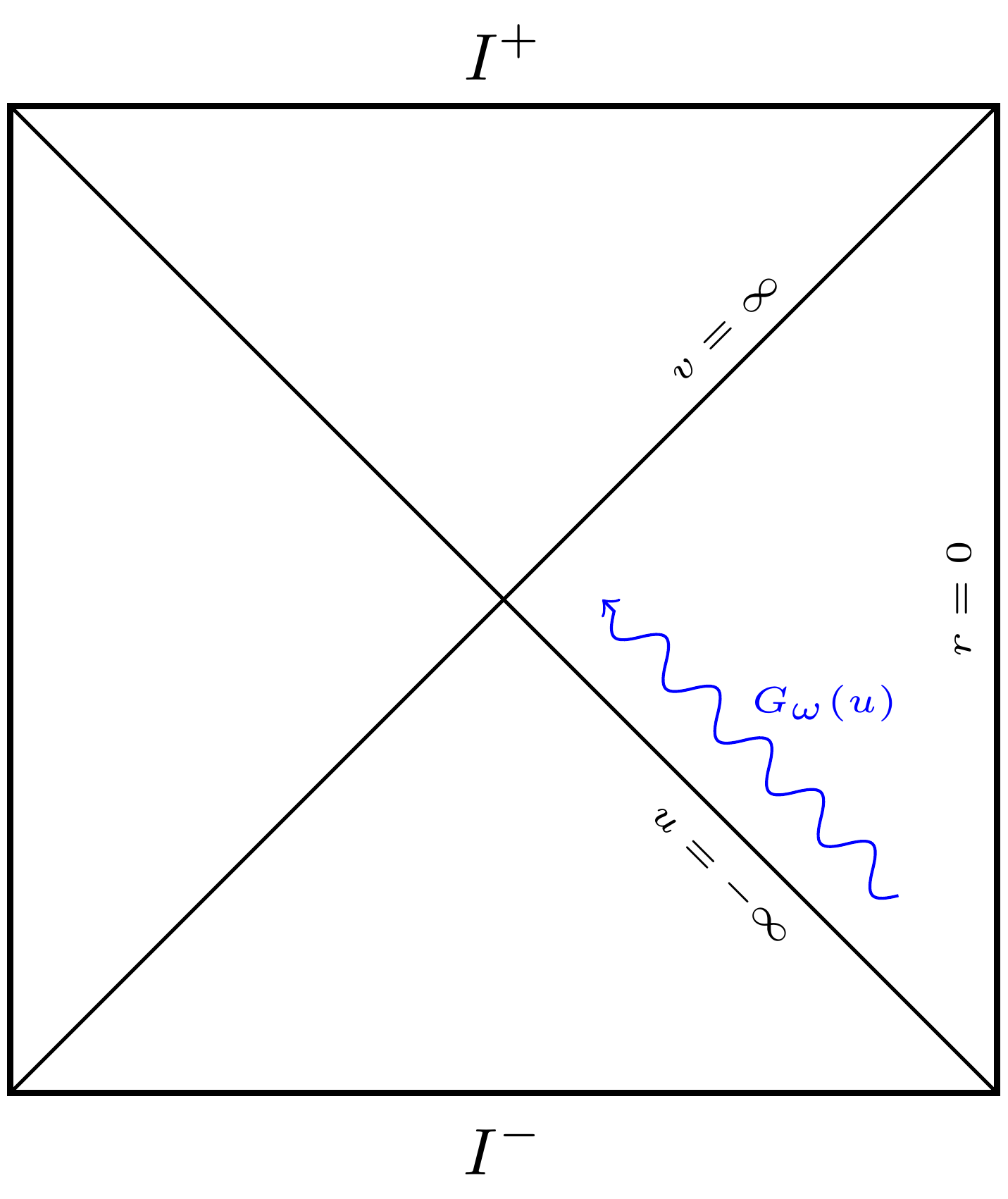} 
	\caption{Space-time diagram of de Sitter space.  Left-moving modes contribute to Gibbons-Hawking radiation near the future horizon (left) and right-moving modes contribute to Gibbons-Hawking radiation near the past horizon (right).  Here, left-moving means that the modes are moving towards the center of the static patch ($r=0$).  Similarly, right-moving means that the modes are moving away from the center of the static patch (not to be confused with the directions of the arrows on the figure).}
	\label{fig:lr_modes}
\end{figure}

In the Eddington-Finklestein and Kruskal coordinate systems, there are two near horizon limits.  The first one, $v \rightarrow \infty$ or $V \rightarrow 0$, is related to the horizon in the far future.  In this case, Gibbons-Hawking radiation can be obtained from the left-moving modes of the field (see Figure \ref{fig:lr_modes}).  The second limit, $u \rightarrow \infty$ or $U \rightarrow 0$, is related to the horizon in the far past.  In this case, Gibbons-Hawking radiation can be obtained from the right-moving modes of the field.  Both limits can be taken independently and yield the same spectrum.

Let us first bring our attention to the left moving modes, which are relevant for Gibbons-Hawking radiation near the future horizon.  The Bogoliubov coefficients for these modes can be derived from the normalized left-moving eigenfunctions $\tilde{G}_\omega(v)$ and $\tilde{H}_\Omega(V)$ of the Laplacian operator in Eddington-Finkelstein and Kruskal coordinates.  These eigenfunctions are given by
\be
\tilde{G}_\omega(v) = \frac{e^{-i \omega v}}{\sqrt{4 \pi w}} \quad , \quad \tilde{H}_\Omega(V) = \frac{e^{-i \Omega V}}{\sqrt{4 \pi \Omega}} \, .
\ee 
We obtain\footnote{Here, we will be using the convention $f \overleftrightarrow{\partial}_x g = f \partial_x g - g \partial_x f$ for the double arrow derivative.}
\begin{align}
\tilde{\alpha}_{\omega \Omega} & = i \int_{-\infty}^0 dV \tilde{G}^*_\omega(v(V)) \overleftrightarrow{\partial}_{V} \tilde{H}_\Omega(V) = \frac{l}{2 \pi} \sqrt{\frac{\omega}{\Omega}} \left( l \Omega\right)^{i l \omega} e^{\pi l \omega /2} \Gamma(-i l \omega) \\
\tilde{\beta}_{\omega \Omega} & = -i \int_{-\infty}^0 dV \tilde{G}_\omega(v(V))  \overleftrightarrow{\partial}_{V} \tilde{H}_\Omega(V) = \frac{l}{2 \pi} \sqrt{\frac{\omega}{\Omega}} \left( l \Omega\right)^{-i l \omega} e^{-\pi l \omega /2} \Gamma(i l \omega) \, ,
\label{eq:betat}
\end{align}
where $\Gamma(z)$ is the gamma function.  Using the Bogoliubov transformations above, we can compute the number of particles an Eddington-Finklestein observer perceives in the Kruskal vacuum.  This number of particles can be obtained from the expectation value of the number operator $\tilde{n}_\omega = \tilde{b}_\omega^\dagger \tilde{b}_\omega$ in the Kruskal vacuum. For an observer near the future horizon, we obtain
\begin{align}
	_a\langle 0 | \tilde{b}_{\omega_1}^\dagger \tilde{b}_{\omega_2} |0\rangle_a & = \int_{0}^{\infty} d\Omega \tilde{\beta}_{\omega_1 \Omega} \tilde{\beta}^*_{\omega_2 \Omega} \\
	& = \frac{1}{e^{2\pi l \omega_1} - 1 } \delta(\omega_1 - \omega_2) \, ,
 \label{eq:spec1}
\end{align}
which is the expected spectrum of Gibbons-Hawking radiation.  Here, we made use of the identity
\be
|e^{- \pi l \omega/2} \Gamma(\pm i l \omega)|^2 = \frac{2\pi}{l \omega} \left( \frac{1}{e^{2\pi l \omega} - 1}\right) \, 
\label{identity}
\ee
to obtain the result above.  A similar spectrum can be obtained from the right-moving modes, which are relevant for Gibbons-Hawking radiation near the past horizon.  In this case, the normalized wave functions are
\be
G_\omega(u) = \frac{e^{-i \omega u}}{\sqrt{4 \pi w}} \quad , \quad H_\Omega(U) = \frac{e^{-i \Omega U}}{\sqrt{4 \pi \Omega}} \, ,
\ee
and the associated Bogoliubov transformations are
\begin{align}
\alpha_{\omega \Omega} & = i \int_{0}^{\infty} dU G^*_\omega(u(U)) \overleftrightarrow{\partial}_{U} H_\Omega(U) = \frac{l}{2 \pi} \sqrt{\frac{\omega}{\Omega}} \left( l \Omega\right)^{-i l \omega} e^{\pi l \omega /2} \Gamma(i l \omega) \\
\beta_{\omega \Omega} & = -i \int_{0}^{\infty} dU G_\omega(u(U))  \overleftrightarrow{\partial}_{U} H_\Omega(U) = \frac{l}{2 \pi} \sqrt{\frac{\omega}{\Omega}} \left( l \Omega\right)^{i l \omega} e^{-\pi l \omega /2} \Gamma(-i l \omega) \, .
\label{eq:beta}
\end{align}
By computing the expectation value of the number operator $n_\omega = b_\omega^\dagger b_\omega$ in the Kruskal vacuum, we obtain the same result as for the left moving modes:
\begin{align}
	_a\langle 0 | b_{\omega_1}^\dagger b_{\omega_2} |0\rangle_a & = \int_{0}^{\infty} d\Omega \beta_{\omega_1 \Omega} \beta^*_{\omega_2 \Omega} \\
	& = \frac{1}{e^{2\pi l \omega_1} - 1 } \delta(\omega_1 - \omega_2) \, .
\label{eq:spec2}
\end{align}
Consequently, at all times, a static observer in de Sitter space will perceive a thermal bath of particles with a temperature given by $T = (2\pi l)^{-1}$.

\subsection{UV cutoff, blue shift and scrambling time}

At first glance, the thermal spectrum of Equation \ref{eq:spec1} and \ref{eq:spec2} does not seem to change in the UV limit.  However, we must remember that physics at low energies is described by an effective field theory and cannot be trusted beyond the Planck scale, where quantum gravity corrections become important.  To take this into account, let us impose a cutoff $\Lambda_{\Omega} \sim M_p$ on the frequency $\Omega$.  When $\Lambda_{\Omega} \gg H$ ($H = l^{-1}$ is the Hubble 
 expansion rate), Equations \ref{eq:spec1} and \ref{eq:spec2} are still approximately valid.  However, there are two regions of the static patches where we expect that new physics might be important.  Given our cutoff, these are the regions near the future and past horizon.  
 
 To see why this is the case, let us study how one particle states in Eddington-Finklestein coordinates behave in the Kruskal frame.  In the Eddington-Finklestein frame, the free particle wave functions are defined by
 \be
 \psi_R(u) = \, _b\langle 0 | \phi(u,v) b^{\dagger}_\omega |0\rangle_b = \frac{e^{-i \omega u}}{\sqrt{4 \pi w}} \quad , \quad \psi_L(v) = \, _b\langle 0 | \phi(u,v) \tilde{b}^{\dagger}_\omega |0\rangle_b = \frac{e^{-i \omega v}}{\sqrt{4 \pi w}} \,
 \ee 
 where $\psi_R(u)$ describes a right moving particle and $\psi_L(v)$ a left moving particle.  In the Eddington-Finklestein frame, $\psi_R(u)$ and $\psi_L(v)$ are eigenfunctions of the momentum operators $p_u = i \frac{d}{du}$ and $p_v = i \frac{d}{dv}$ with an eigenvalue given by the frequency $\omega$ associated to each wave function.  Similarly, $\psi_R(u)$ and $\psi_L(v)$ are also eigenfunctions of the momentum operators $p_U = i \frac{d}{dU}$ and $p_V = i \frac{d}{dV}$ in Kruskal coordinates.  Acting with these operators, we obtain
 \be
i \frac{d \psi_R(u)}{d U} = \omega \frac{du}{dU} \psi_R(u) \quad , \quad i \frac{d \psi_L(v)}{d V} = \omega \frac{dv}{dV} \psi_L(v) \, ,
 \ee
from which we read off the frequency eigenvalues 
 \be
\Omega = \frac{dv}{dV}\omega = e^{v/l} \omega \quad , \quad \Omega = \frac{du}{dU} \omega = e^{-u/l}\omega\, .
\ee
From the results above, we conclude that the Kruskal frame frequency associated to Eddington-Finklestein particles is blue-shifted close the future and past horizon ($v \rightarrow \infty$ and $u \rightarrow - \infty$ limits).  

In these, limits we have to worry about this frequency being blueshifted above our UV cutoff $\Lambda_{\Omega} \sim M_p$.  As an example, let us consider Eddington-Finklestein particles with frequencies on the Hubble scale ($\omega \sim H$), which are those that contribute most significantly to Gibbons-Hawking radiation.  For such particles, the blueshifted frequency exceeds the UV cutoff when $v \gg l \ln (l M_p )$ for a left-moving particle or $u \ll - l \ln (l M_p )$ for a right-moving particle.  For an observer at the center of the static patch ($r=0$), these bounds correspond precisely to a scrambling time $t_{sc} = l \ln (l M_p )$ in the future and in the past.  After this amount of time, we expect UV effects to become relevant in Gibbons-Hawking radiation.

From what we have seen so far, the Gibbons-Hawking spectrum coming from particles that behave as plane waves does not seem sensitive to the effects above.  We will see that this is a result of particles being described as plane waves as opposed to localized wave packets.  The key difference between the two descriptions is that wave-packets are localised within a left or right-moving patch of width $\Delta v$ or $\Delta u$ depending on the direction of motion of the wave packet.  In comparison, plane waves are fully non-local and span the whole static patch at all times.  Because the wave-packets are local, they are sensitive to the region of the static patch where they propagate.  Hence, they will be sensitive to the UV cutoff in the blue-shifted regions $v \gg l \ln (l M_p )$ and $u \ll - l \ln (l M_p )$.  In the next sections, we will see that this sensitivity leads to a shutdown of Gibbons-Hawking radiation when $v \gg l \ln (l M_p )$ or $u \ll - l \ln (l M_p )$.

%This difference will prove important for the behavior of Gibbons-Hawking radiation near the future and past horizon of de Sitter.

\section{Wave packets in de Sitter space}
\label{sec:wp}

Let us now turn our attention to the case where particles are described by semi-classical wave packets.  We will be interested in left-moving wave packets $\psi_{(w_0 , v_0)}(v)$ and right-moving wave packets $\psi_{(w_0 , u_0)}(u)$ of the form
\be
\psi_{(w_0 , v_0)}(v) = \int_{0}^{\infty} \frac{d\omega}{\sqrt{4 \pi w}} f_{w_0}(\omega) e^{- i \omega (v-v_0)} \quad , \quad \psi_{(w_0 , u_0)}(u) = \int_{0}^{\infty} \frac{d\omega}{\sqrt{4 \pi w}} f_{w_0}(\omega) e^{- i \omega (u-u_0)} \, .
\label{eq:wave_f}
\ee
Here, we will consider frequency distributions $f_{\omega_0}(\omega)$ that are peaked around a central frequency $\omega_0$, and vanishing a distance $\Delta \omega$ from $\omega_0$.  This way, the left and right moving wave packets will become centered around $v_0$ and $u_0$ respectively.  For the wave packets to be properly normalized, we will impose the normalisation condition 
\be
\int_{-\infty}^{\infty} dv \rho_{(\omega_0 , v_0)}(v) = \int_{0}^{\infty} d\omega |f_{\omega_0}(\omega)|^2 = 1 \, ,
\ee
where 
\be
\rho_{(\omega_0 , u_0)}(v) = i \psi^*_{(w_0 , v_0)} \overleftrightarrow{\partial}_v \psi_{(w_0 , v_0)}
\label{eq:density}
\ee
is the relativistic density of the wave packet $\psi_{(w_0 , v_0)}(v)$. (Here, we used the left-moving wave packet for this definition, but a similar condition is true for the right-moving wave packet as well.)  One of the most simple frequency distributions which satisfies the conditions above is the Gaussian distribution
\be
f_{\omega_0}(\omega) = \sqrt{\frac{\omega}{2\pi \omega_0 \Delta \omega}} e^{- \frac{(\omega - \omega_0)^2}{2 \Delta \omega^2}} \, ,
\label{eq:freq_dist}
\ee
associated to the left and right-moving Gaussian wave packets
\be
\psi_{(w_0 , v_0)}(v) \approx \sqrt{\frac{\Delta \omega}{2\sqrt{\pi} \omega_0} }e^{- \frac{\Delta \omega^2 (v-v_0)}{2} - i \omega_0(v-v_0)} \quad , \quad \psi_{(w_0 , u_0)}(u) \approx \sqrt{\frac{\Delta \omega}{2\sqrt{\pi} \omega_0} }e^{- \frac{\Delta \omega^2 (u-u_0)}{2} - i \omega_0(u-u_0)} \, ,
\label{eq:wave_p}
\ee
in the limit where $\Delta \omega \ll \omega_0$.  The wave functions above saturate the Heisenberg uncertainty bound.  Hence, we should view particles described by $\psi_{(w_0 , v_0)}(v)$ and $\psi_{(w_0 , u_0)}(u)$ as free particles in their semi-classical limits. 

Such particles can be created in quantum field theory by appropriately modifying the raising and lowering operators.  Using $b_\omega^\dagger , \tilde{b}_\omega^\dagger$ and the frequency distribution $f_{\omega_0}(\omega)$ of Equation \ref{eq:freq_dist}, we can define creation operators $\tilde{b}^\dagger_{(\omega_0 , v_0)} , b^\dagger_{(\omega_0 , u_0)}$ associated to the Gaussian wave packets $\psi_{(w_0 , v_0)}(v)$ and $\psi_{(w_0 , u_0)}(u)$.  Such creation operators are given by
\be
\tilde{b}^\dagger_{(\omega_0 , v_0)} = \int_0^\infty d\omega f_{\omega_0}(\omega) e^{i \omega v_0} \tilde{b}^{\dagger}_{\omega} \quad , \quad  b^\dagger_{(\omega_0 , u_0)} = \int_0^\infty d\omega f_{\omega_0}(\omega) e^{i \omega u_0} b^{\dagger}_{\omega} \, .
\label{eq:raising}
\ee
Similar annihilation operators can be obtained by taking the hermitian conjugate of the operators above.  One can check that this creation operator lets us create particles with the wave function of Equation \ref{eq:wave_f} by evaluating the wave function of a one-particle state in quantum field theory.  For the left-moving particle, the wave function can be obtained by evaluating the vacuum expectation value of $\phi(v) \tilde{b}^{\dagger}_{(\omega_0 , v_0)}$ in the Eddington-Finkelstein vacuum.  Similarly, the wave function of the right moving particle can be obtained by computing the Eddington-Finkelstein vacuum expectation value of $\phi(u) b^{\dagger}_{(\omega_0 , u_0)}$.  This gives us
\be
_b\langle 0 | \phi(v) \tilde{b}^\dagger_{(\omega_0,v_0)} | 0 \rangle_b = \psi_{(\omega_0 , v_0)}(v)  \quad , \quad _b\langle 0 | \phi(u) b^\dagger_{(\omega_0,u_0)} | 0 \rangle_b = \psi_{(\omega_0 , u_0)}(u)\, ,
\ee
as expected.  The fully quantum limit of this semi-classical particle can be recovered by letting the $\Delta \omega \rightarrow 0$.  In this case, we recover the quantum mechanical wave function of a free particle that behaves as a plane wave\footnote{It's a bit tricky to see this from Equation \ref{eq:wave_p}.  However, we can see that this is true when taking the $\Delta \omega \rightarrow 0$ limit in Equation \ref{eq:freq_dist}.  In this limit, we obtain $f_{\omega_0}(\omega) \approx \sqrt{\omega/\omega_0} \delta(w - w_0)$.  Evaluating Equation \ref{eq:wave_f} using this frequency distribution, we obtain the desired result.}:
\be
\psi_{(w_0 , v_0)}(v) \approx  \frac{e^{- i \omega_0(v-v_0)}}{\sqrt{4 \pi \omega_0}} \quad , \quad \psi_{(w_0 , u_0)}(u) \approx \frac{e^{- i \omega_0(u-u_0)}}{\sqrt{4 \pi \omega_0}}\, .
\ee
Using $\tilde{b}^\dagger_{(\omega_0 , v_0)} , b^\dagger_{(\omega_0 , u_0)}$ and the corresponding annihilation operators, it is possible to compute the spectrum of Gibbons-Hawking radiation associated to a thermal bath of semi-classical particles in the Kruskal vacuum.  This can be done by defining number operators $\tilde{\mathcal{N}}_{(\omega_0 , v_0)} = \tilde{b}^\dagger_{(\omega_0 , v_0)}\tilde{b}_{(\omega_0 , v_0)}$ and $\mathcal{N}_{(\omega_0 , u_0)} = b^\dagger_{(\omega_0 , u_0)} b_{(\omega_0 , u_0)}$ associated to the left and right-moving wave packets, and computing their expectation values in the Kruskal vacuum in the same way we did in section \ref{sec:rev}.  As we will see in the next section, the spectrum we obtain will be similar to our previous result, but also sensitive to UV cutoffs.

\section{Shutdown of Gibbons-Hawking radiation in de Sitter space}
\label{sec:shut}

We now provide a derivation of Gibbons-Hawking radiation in de Sitter space in the case where particles are described as wave packets rather than plane waves.  In this case, the Gibbons-Hawking spectrum will vary depending on the location $v_0$ or $u_0$ of the contributing wave packets.  When $v_0 \ll l \ln (l M_p)$ and $u_0 \gg - l \ln (l M_p)$, the spectrum will remain approximately the same as in equation \ref{eq:spec1} and \ref{eq:spec2}.  Conversely, the spectrum will shut down in the cases where $v_0 \gg l \ln (l M_p)$ and $u_0 \ll - l \ln (l M_p)$.  Since Gibbons-Hawking radiation becomes sensitive to the UV cutoff with the localized wave packet formalism, $v_0 \approx l \ln (l M_p)$ and $u_0 \approx - l \ln (l M_p)$ give bounds on the validity of effective field theory near the future and past horizons.  We will see that these bounds match the predictions from the TCC for an observer at the center of the static patch ($r_* = 0$).
%Indeed, near these horizons, the wave packets are blueshifted and Planckian effects are present.  

\subsection{Late-time and early-time spectrum}

Following the same steps as in section \ref{sec:rev}, let us compute the expectation value of the number operators $\tilde{\mathcal{N}}_{(\omega_0 , v_0)}$ and $\mathcal{N}_{(\omega_0 , u_0)}$ in the Kruskal vacuum. The first step is to substitute the raising operators of equation \ref{eq:raising} and the corresponding annihiliation operators in the definition of the number operators.  We obtain the following expectation values:
\begin{align}
_a\bra 0  \tilde{\mathcal{N}}_{(\omega_0 , v_0)} \ket 0_a  & = \int_{0}^{\infty} d\omega_{1}\int_{0}^{\infty}d\omega_{2} f_{\omega_{0}}(\omega_{1})f^{*}_{\omega_{0}}(\omega_{2})e^{i(\omega_{1}-\omega_{2})v_{0}}\int_{0}^{\Lambda_{\Omega}}d\Omega\tilde{\beta}_{\omega_{1} \Omega}^{*}\tilde{\beta}_{\omega_{2} \Omega} \, , \\
_a\bra 0  \mathcal{N}_{(\omega_0 , u_0)} \ket 0_a & = \int_{0}^{\infty} d\omega_{1}\int_{0}^{\infty}d\omega_{2} f_{\omega_{0}}(\omega_{1})f^{*}_{\omega_{0}}(\omega_{2})e^{i(\omega_{1}-\omega_{2})u_{0}}\int_{0}^{\Lambda_{\Omega}}d\Omega\beta_{\omega_{1} \Omega}^{*}\beta_{\omega_{2} \Omega} \, .
\end{align}
Here, we have imposed the UV cutoff $\Lambda_\Omega = M_p$ on the energy scale $\Omega$ in order to remove UV effects.  Moreover, $\beta_{\omega_{} \Omega}$ and $\tilde{\beta}_{\omega_{} \Omega}$ are given by \ref{eq:beta} and \ref{eq:betat} respectively.
%With (\ref{eq:beta}) formula for $\beta_{\omega_{} \Omega}$ and with (\ref{eq:betat}) formula for $\tilde{\beta}_{\omega_{} \Omega}$.\\
To see the impacts of the UV cutoff on the contribution of wave packets in different regions of the static patch, we will then make the change of variables $\Omega= \omega_0 e^{v/l}$ for the left-moving modes and $\Omega= \omega_0e^{-u/l}$ for the right-moving modes.  These correspond to the blue-shifted frequency of left or right-moving wave packet of central frequency $\omega_0$.  Following the change of variables, the expectation value of the number operators becomes
\begin{align}
 _a \bra 0  \tilde{\mathcal{N}}_{(\omega_0 , v_0)} \ket 0 _a & = \int_{-\infty}^{l \log(\Lambda_{\Omega}/\omega_0)}
 \dfrac{dv}{l}\lvert \tilde{F}_{\omega_{0}}(v-v_{0}) \rvert^{2} \, , \\
 _a\bra 0  \mathcal{N}_{(\omega_0 , u_0)} \ket 0 _a & = \int_{- l \log( \Lambda_{\Omega}/\omega_0)}^{\infty}
 \dfrac{du}{l}\lvert F_{\omega_{0}}(u-u_{0}) \rvert^{2} \, ,
\end{align}
where $\tilde{F}_{\omega_{0}}$ and $F_{\omega_{0}}$ are given by
\begin{align}
\tilde{F}_{\omega_{0}}(v-v_{0}) &= \dfrac{l}{2\pi} \int_{0}^{\infty} d\omega f_{\omega_{0}}(\omega)e^{-i\omega(v-v'_{0})}  \sqrt{\omega}  e^{-\pi  \omega l/2}  \Gamma(i l \omega) \, , \\
F_{\omega_{0}}(u-u_{0}) &= \dfrac{l}{2\pi} \int_{0}^{\infty} d\omega f_{\omega_{0}}(\omega)e^{-i\omega(u-u'_{0})}  \sqrt{\omega}  e^{-\pi  \omega l/2}  \Gamma(-i l \omega) \, .
\end{align}
Here, the centers $v'_0$ and $u'_0$ for each distribution are given by $v'_0 = v_0 - l \ln (l \omega_0)$ and $u'_0 = u_0 + l \ln (l \omega_0)$. When the frequency distribution $f_{\omega_0}(\omega)$ is sufficiently narrow, it is possible to approximate the expressions above as
%With a Taylor expansion around $\omega_{0}$, $F$ can be written as :
\begin{align}
\tilde{F}_{\omega_{0}}(v-v_{0}) & \approx \dfrac{l}{\sqrt{\pi}}\omega_{0}e^{-\pi  \omega_{0}l/2}\Gamma(i l \omega_{0})\psi_{\omega_{0},v'_{0}}(v) \, ,\\ 
\tilde{F}_{\omega_{0}}(u-u_{0}) & \approx \dfrac{l}{\sqrt{\pi}}\omega_{0}e^{-\pi  \omega_{0}l/2}\Gamma(-i l \omega_{0})\psi_{\omega_{0},u'_{0}}(u) \, .
\end{align}
For frequencies on cosmic scales $\omega_0 \sim l$, the approximation above holds if $\Delta \omega \ll \omega_0$. Using the expressions above and the identity (\ref{identity}), we can finally express $_a\bra 0  \tilde{\mathcal{N}}_{(\omega_0 , v'_0)} \ket 0_a$ and $_a\bra 0  \mathcal{N}_{(\omega_0 , u'_0)} \ket 0_a$ as
\begin{align}
_a\bra 0 \tilde{\mathcal{N}}_{(\omega_0 , v_0)} \ket 0_a & \approx \frac{1}{e^{2\pi l \omega_0} - 1} \left[\int_{-\infty}^{l \ln(\Lambda_{\Omega}/\omega_0)} dv \rho_{(\omega_0 , v'_0)}(v)\right] \, , 
\label{eq:wave_p_spec1}\\
_a\bra 0 \mathcal{N}_{(\omega_0 , u_0)} \ket 0_a & \approx \frac{1}{e^{2\pi l \omega_0} - 1} \left[\int_{-l \ln(l \Lambda_{\Omega}/\omega_0)}^{\infty} du \rho_{(\omega_0 , u'_0)}(u)\right] \, ,
\label{eq:wave_p_spec2}
\end{align}
where $\rho_{(\omega_0 , v'_0)}(v)$ and $\rho_{(\omega_0 , u'_0)}(u)$ are the relativistic density of the wave-packet $\psi(\omega_{0},v'_{0})(v)$ and $\psi(\omega_{0},u'_{0})(u)$ (see Equation \ref{eq:density}).  As we can see from equations \ref{eq:wave_p_spec1} and \ref{eq:wave_p_spec2}, wave packets above the cosmic scale $\omega_0 \sim l$ are exponentially suppressed.  Since only wave-packets of frequency $\omega_0 \sim l$ and below contribute significantly to the spectrum, we will be assuming $\omega_0 \sim l$ for the rest of the computations.  For such frequencies, we have $v'_0 = v_0$ , $u'_0 = u_0$ and the bounds of integration in equation \ref{eq:wave_p_spec1} and \ref{eq:wave_p_spec2} correspond to a scrambling time $v_{sc} = l \ln(l M_p)$ in the future and $u_{sc} = -l \ln (l M_p)$ in the past.

\subsection{Early and late time bounds}

\begin{figure}[h]
	\centering
        \includegraphics[width=6cm]{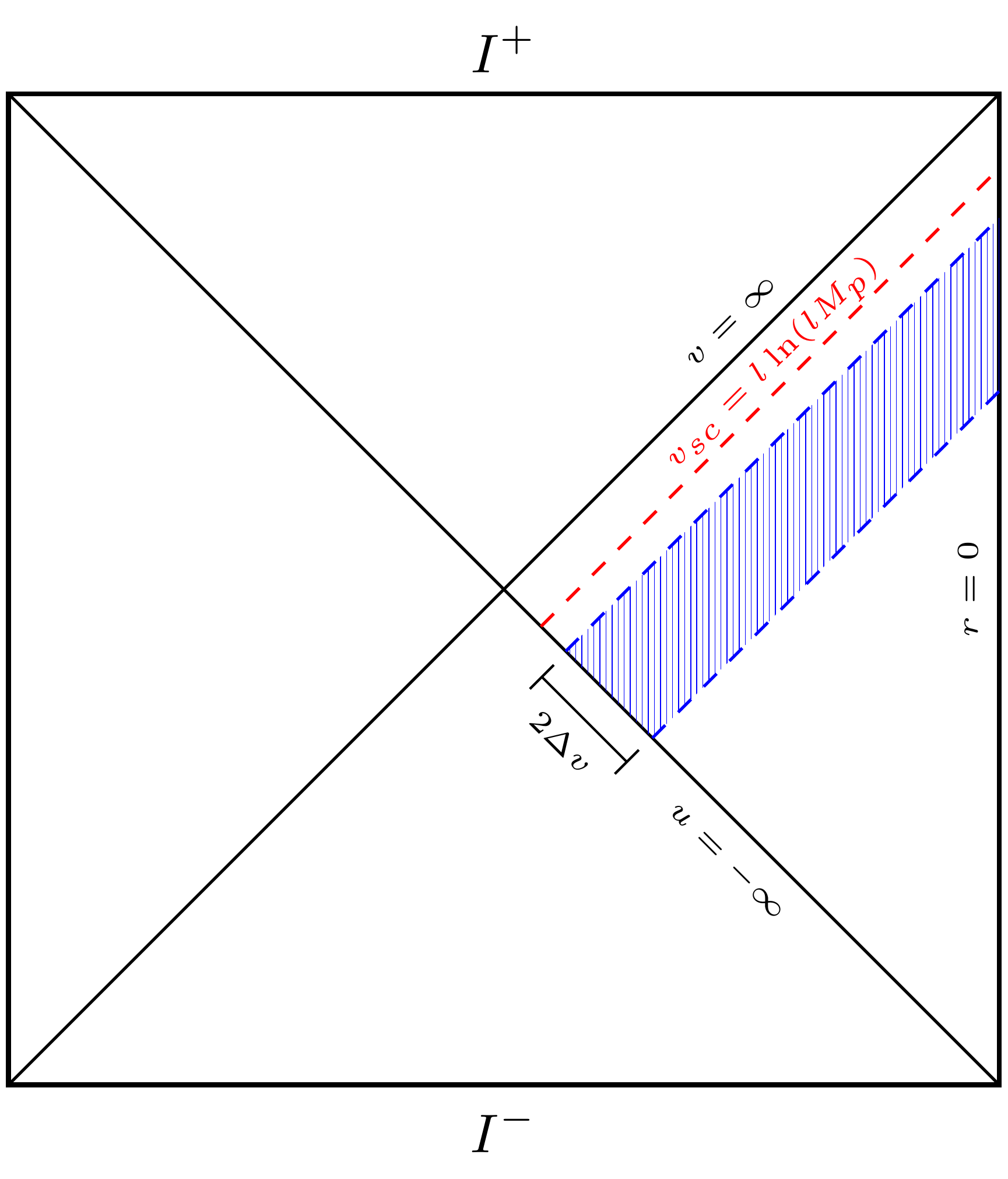} \hspace{1cm}
        \includegraphics[width=6cm]{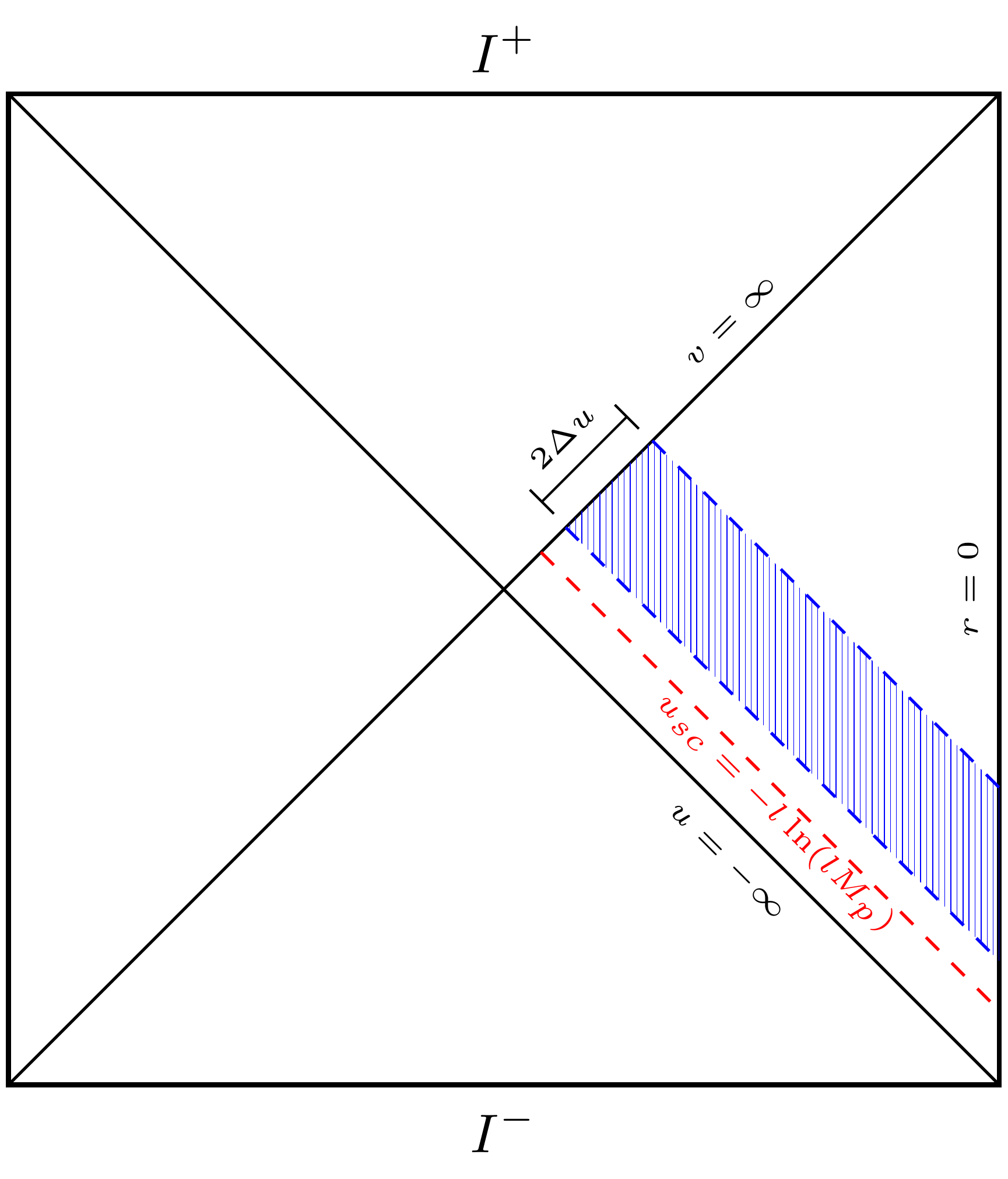} 
	\caption{Left and right-moving wave packets are peaked in a light-cone region of width $2 \Delta v$ (left figure) centered around $v_0$ or a light-cone region of width $2 \Delta u$ centered around $u_0$ (right figure) depending on the direction of motion.  The expectation values of the number operators give us the Gibbons-Hawking spectrum as long as the light cone regions are found below $v_{sc} = l \ln (l M_p)$ or above $u_{sc} = -l \ln (l M_p)$.  When the regions are found above the $v_{sc} = l \ln (l M_p)$ or below $u_{sc} = -l \ln (l M_p)$, the expectation values of the number operators become zero.}
    \label{fig:le_bounds}
\end{figure}

The number operators of equation \ref{eq:wave_p_spec1} and \ref{eq:wave_p_spec2} are simply the Planck distribution multiplied by the probabilities $\int_{-\infty}^{v_{sc}} dv \rho_{(\omega_0 , v_0)}(v)$ , $\int_{u_{sc}}^{\infty} du \rho_{(\omega_0 , u_0)}(u)$ of finding the particle described by the wave packet in the intervals $v\in \left]-\infty, v_{sc} \right]$ and $u \in \left[u_{sc},\infty \right[$.  Here, the wave-packets are found in the regions $[ v_0 - \Delta v , v_0 + \Delta v]$ , $[ u_0 - \Delta u , u_0 + \Delta u]$, where $\Delta v$ and $\Delta u$ are related to $\Delta \omega$ via $\Delta u = \Delta v = 1/\Delta \omega$.  When $v_{sc} \geq v_{0}+\Delta v$ or $u_{sc} \leq u_{0}-\Delta u$, the wave packets will almost always be found within the bounds of integration.  In this case, the probabilities give us
\be
\int_{-\infty}^{v_{sc}} dv \rho_{(\omega_0 , v_0)}(v) \approx 1 \quad , \quad \int_{u_{sc}}^{\infty} du \rho_{(\omega_0 , u_0)}(u) \approx 1 \ ,
\label{eq:condition}
\ee
and we recover usual Gibbons-Hawking spectrum. Conversely, when $v_{0}-\Delta v \geq v_{sc}$ or $u_{0}+\Delta u \leq u_{sc}$, the wave packets will almost always be found in the regions forbidden by the cutoff $\Lambda_\Omega$, which are outside the bounds of integration.  In this case, the probabilities give us 
\be
\int_{-\infty}^{v_{sc}} dv \rho_{(\omega_0 , v_0)}(v) \approx 0 \quad , \quad \int_{u_{sc}}^{\infty} du \rho_{(\omega_0 , u_0)} (u) \approx 0 \, ,
\ee
and the expectation values of the number operators become zero (see Figure \ref{fig:le_bounds}).  This shutdown can be illustrated by using the Gaussian wave packets of equation \ref{eq:wave_p} as an example.  In this case, the wave packet densities are given by 
\be
\rho_{(\omega_0 , v_0)}(v) = \frac{1}{\sqrt{\pi} \Delta v} e^{-\frac{(v-v_0)^2}{\Delta v^2}} \quad , \quad \rho_{(\omega_0 , v_0)}(u) = \frac{1}{\sqrt{\pi} \Delta u} e^{-\frac{(u-u_0)^2}{\Delta u^2}}
\ee
From these densities, we obtain the following expectation values for the number operators.
\begin{align}
_a\bra 0 \tilde{\mathcal{N}}_{(\omega_0 , v_0)} \ket 0_a & = \frac{1}{2} \frac{1}{e^{2\pi l \omega_0} - 1} \text{erfc}\left(\frac{v_0 - v_{sc}}{\Delta v}\right) \\
_a\bra 0 \mathcal{N}_{(\omega_0 , u_0)} \ket 0_a & = \frac{1}{2} \frac{1}{e^{2\pi l \omega_0} - 1} \text{erfc}\left(- \frac{u_0 - u_{sc}}{\Delta u}\right)\, .
\end{align}
Here, erfc$(x)$ is the complementary error function.  As expected, we recover the Gibbons-Hawking spectrum when $v_{sc} \geq v_{0}+\Delta v$ or $u_{sc} \leq u_{0}-\Delta u$.  When $v_{0}-\Delta v \geq v_{sc}$ or $u_{0}+\Delta u \leq u_{sc}$, the spectrum becomes exponentially suppressed.  Eventually, the spectrum becomes zero when $v_0 \gg v_{sc}$ and $u_0 \ll u_{sc}$.

From the wave packet example, we can also see that the shut down is truly an effect that arises when particles are localised in space.  In the limits $\Delta v \rightarrow \infty$ and $\Delta u \rightarrow \infty$, where particles behave as plane waves, some particles will always be found the region allowed by the UV cutoff, and the complementary error function erfc$(x)$ will always be equal to one. In this case, we recover the Gibbons-Hawking spectrum of equations \ref{eq:spec1} and \ref{eq:spec2} up to an overall factor of 1/2.  This extra factor of 1/2 arises as a consequence of imposing the UV cutoff.  If we take the limit $M_p \rightarrow \infty$ in a way that $v_{sc} \geq v_{0}+\Delta v$ and $u_{sc} \leq u_{0}-\Delta u$ are satisfied, then equations \ref{eq:spec1} and \ref{eq:spec2} are fully recovered.

\begin{figure}[t]
	\centering
        \includegraphics[width=8.1cm]{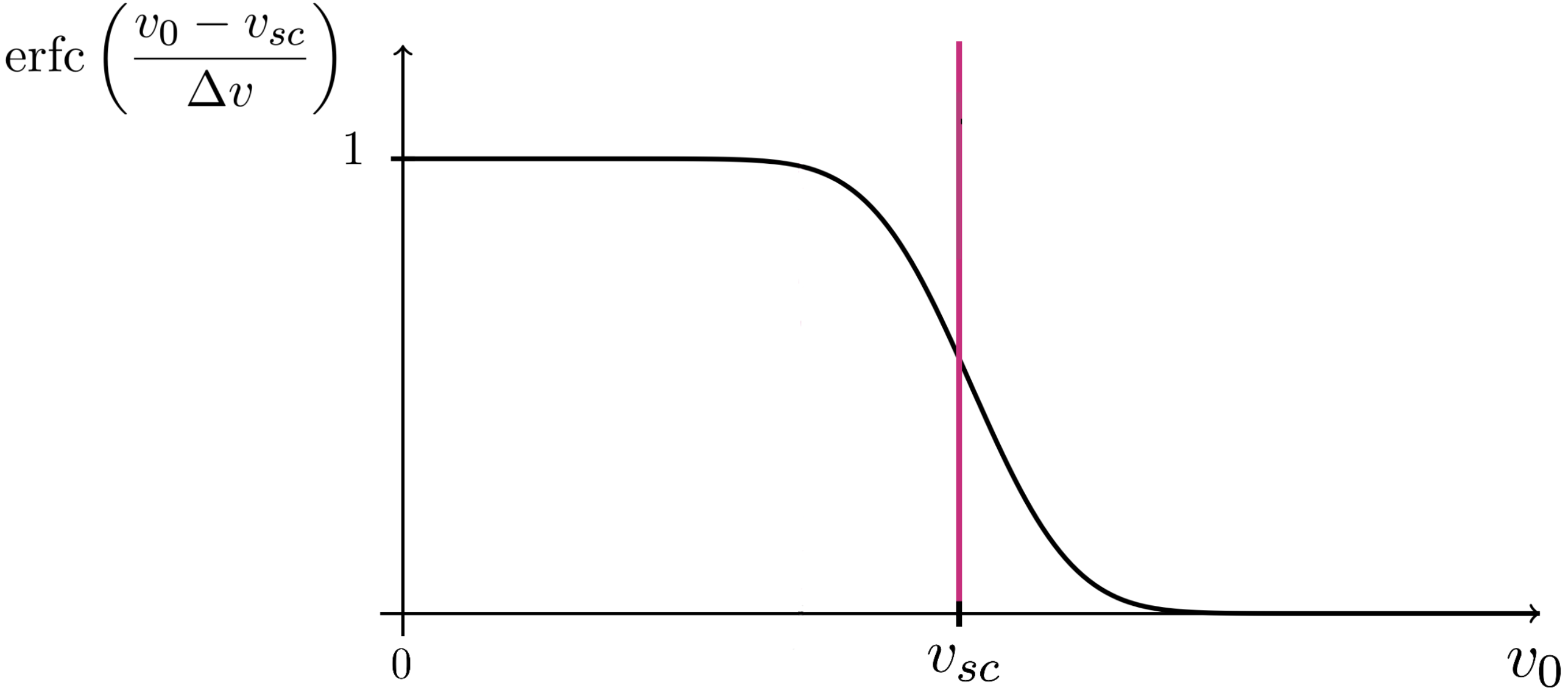} \hspace{0.1cm}
        \includegraphics[width=8.1cm]{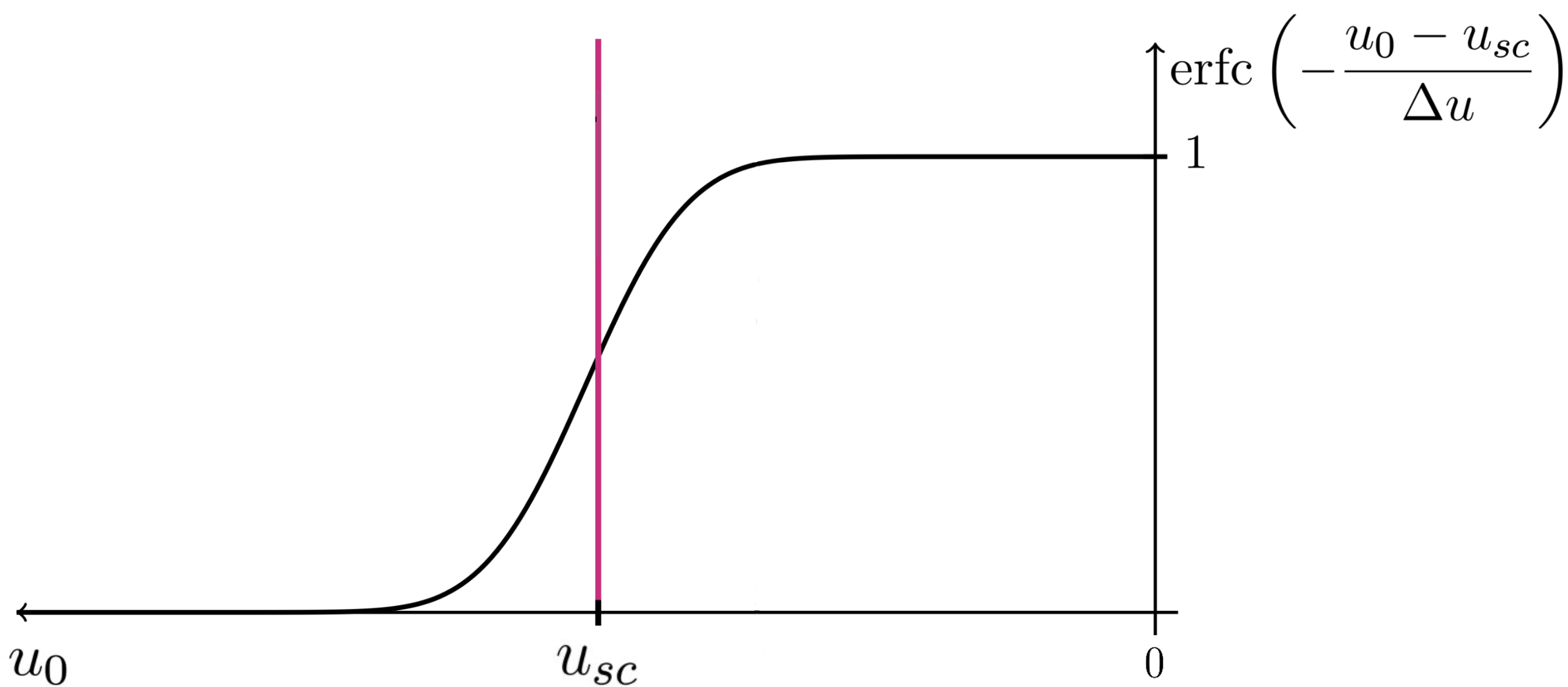} 
	\caption{Illustration of the shutdown in the Gaussian wave packets case for the late time bound on the left and the early time bound on the right. As soon as the conditions $v_{0}+\Delta v \leq v_{sc}$ and $u_{sc} \leq u_{0}-\Delta u$ are no longer satisfied, the Gaussian wave packet finds itself in the region forbidden by the UV cutoff and the spectrum becomes exponentially suppressed due to the cutoff. }
    \label{uvwavepacket}
\end{figure}

%This example illustrates the impact of the UV cutoff on the  Gibbons-Hawking radiation. %This spectrum is present and constant but from a certain $v_{sc}$ and $u_{sc}$, it is exponentially supressed. 
%As shown in the previous section and with the example of the Gaussian wave packet, the radiation is quickly shutdown from $v_{sr}=\log(l\Lambda_{\Omega})l$ and $u_{sr}=- l \log(l\Lambda_{\Omega})$ as a result of imposing the UV cutoff. %This computation allows to confirm a late time bound on the duration of the Sitter space as an effective field theory and corresponds to the predictions of the Trans-Planckian censorship conjecture.

%Approximation:
%\be
%\int_{v_0-\Delta v}^{v_0+\Delta v} dv \rho_{(\omega_0 , v_0)}(v) \approx 1
%\ee

\subsection{Relation to the TCC bound}

\begin{figure}[h]
	\centering
	\includegraphics[width=7cm]{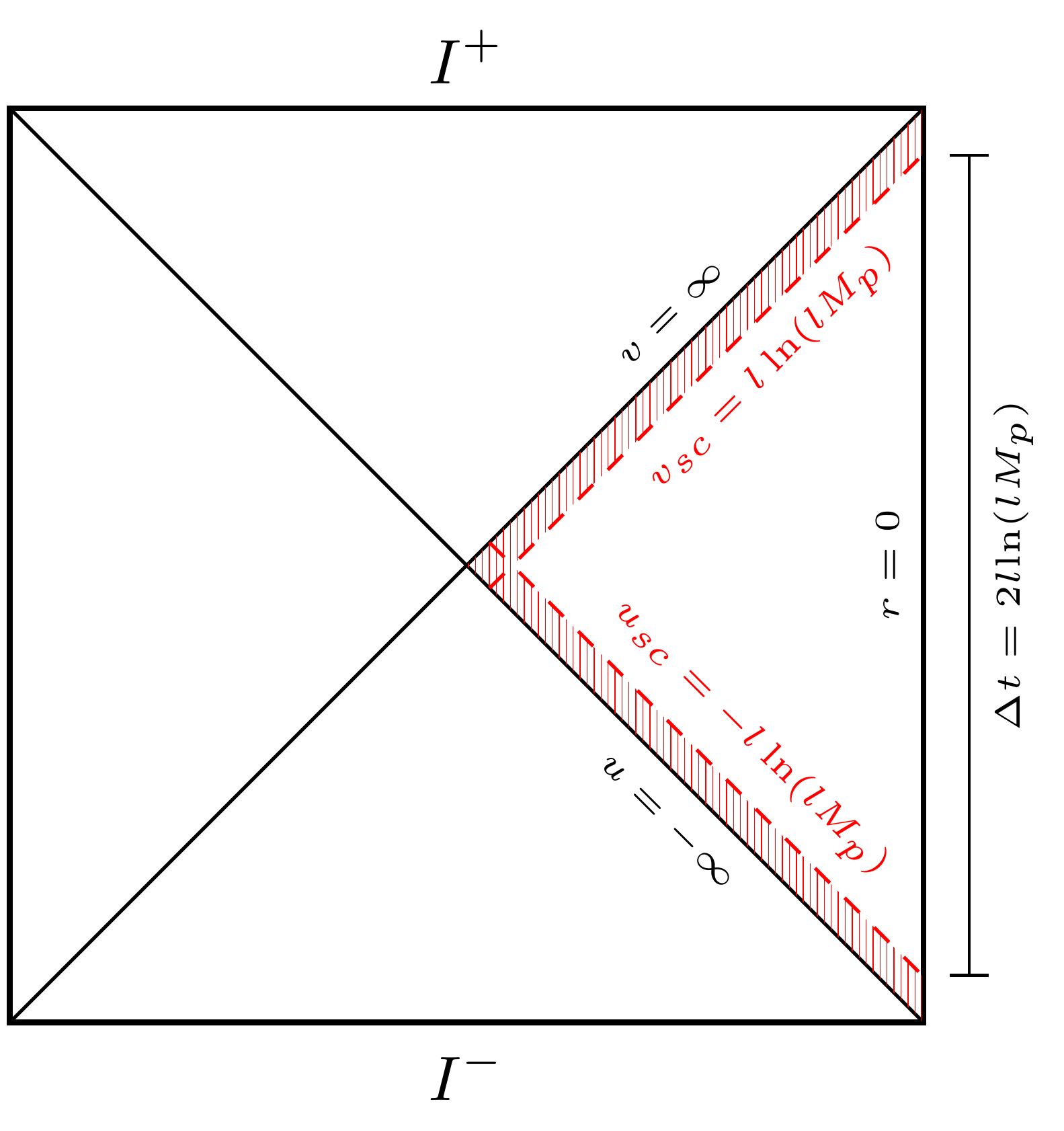}
	\caption{Assuming effective field theory is valid for when $u > -l \ln (l M_p)$ and $v < l \ln (l M_p)$, we obtain a bound $|t| > l \ln (l M_p)$ on the validity of effective field theory for an observer at the center of the static patch ($r_* = 0$).  This region corresponds to time interval $\Delta t = 2 l \ln (l M_p)$ centered around $t=0$.}
	\label{fig:c_bounds}
\end{figure}

%The example of the previous section illustrates the impact of the UV cutoff on Gibbons-Hawking radiation when particles are described as wave packets. %This spectrum is present and constant but from a certain $v_{sc}$ and $u_{sc}$, it is exponentially supressed. 
%As shown in the previous section and with the example of the Gaussian wave packet, 

%the radiation is quickly shutdown from $v_{sr}=\log(l\Lambda_{\Omega})l$ and $u_{sr}=- l \log(l\Lambda_{\Omega})$ as a result of imposing the UV cutoff. %This computation allows to confirm a late time bound on the duration of the Sitter space as an effective field theory and corresponds to the predictions of the Trans-Planckian censorship conjecture.

As we have seen in the previous section, the spectrum of Gibbons-Hawking radiation is shutdown in the regions where $v > v_{sc}$ and $u < u_{sc}$ when particles are described as wave packets.  Since the shutdown arises as a consequence of imposing a UV cutoff, one should not necessary view it as a physical effect, but rather an indication that new physics must be taken into account in the regions where the cutoff is in effect.  Using this interpretation, we conclude that effective field theory must breakdown in the regions $v > v_{sc}$ and $u < u_{sc}$.  

If we assume effective field theory stops being valid when $v$ is above $v_{sc}$ and $u$ is below $u_{sc}$, it is possible to recover a bound on the time scale of validity of effective field theory in de Sitter space.  In terms of satic patch time t, the bounds $v < v_{cs}$ and $u > u_{cs}$ can be expressed as $u_{sc} + r_* < t < v_{sc} - r_*$.  For a static observer sitting at the center of the static patch ($r_* = 0$), this implies $|t| > l \ln (l M_p)$ (See Fibure \ref{fig:c_bounds}).  In other words, the effective field theory description will only be valid for a time interval $\Delta t = 2 l \ln (l M_p)$ centered around $t = 0$, which agrees with the time-scale $\Delta t \sim l \ln (l M_p )$ predicted by the TCC.  In fact, it is exactly twice the TCC time scale.

%which is exactly twice the amount of time predicted by the TCC.  

The reason why we are obtaining twice the TCC time-scale seems to be because we worked under the assumption that space-time is eternally in a de Sitter phase.  In this case, we have to consider bounds in the far future and the far past, which contribute to the time scale by an amount $\Delta t = l \ln (l M_p)$ each.  In realistic cosmologies, a matter or radiation dominated phase will precede the de Sitter phase.  If we impose that the de Sitter phase begins at $t = 0$, then the future bound $v < v_{sc}$ imposes the constraint $t < l \ln (l M_p)$ at $r_* = 0$.  In this case, we recover $\Delta t = l \ln (l M_p)$, which is exactly the amount of time predicted by the TCC.  A similar argument can be made to constrain eternal inflation.  If we assume eternal inflation ends at $t = 0$, then the past bound $u > u_{sc}$ imposes the constraint $t > - l \ln (l M_p)$ at $r_* = 0$.  In this case, we also recover $\Delta t = l \ln (l M_p)$, which is exactly the amount of time predicted by the TCC.

\section{Conclusion and discussion}

We have computed the spectrum of Gibbons-Hawking radiation in the static patch of de Sitter space using two approaches.  First, we used the standard effective field theory approach focusing on the propagation of Fourier modes. Second, we considered the propagation of particles described as wave packets.  For this description, we found that the radiation shuts off beyond the TCC time scale if there is a fundamental ultraviolet cutoff. In this case, the spectrum of this radiation cannot be computed reliably using the usual effective field theory techniques beyond the TCC time.\footnote{Here, it should be stressed that the key difference between our analysis and standard approaches is that we impose that particles are described by wave packets, and our effective field theory is only reliable below an energy cutoff $\Lambda$.  In the limit where the energy cutoff $\Lambda$ and the width of the wave packet go to infinity (plane-wave limit), our analysis shows that there are no issues on the TCC time scale.  In this case, the classical description of de Sitter space may receive quantum corrections at a later time \cite{dSislands}, \cite{Zell}.}

Our result supports the Trans-Planckian Censorship Conjecture \cite{Bedroya1} which implies that the effective field theory of fluctuations about an inflationary background cosmology will break down beyond the TCC time scale \cite{Bedroya2}.

Our analysis is based on applying the techniques used for black holes in \cite{HK} to the case of cosmology.  In the case of black holes, the effective field theory of fluctuations breaks down beyond the scrambling time.  Our analysis points out an interesting analogy between the scrambling time for black holes \cite{scrambling} and the TCC time scale for an inflationary cosmology.  %{ \color{red} Interestingly enough, this analogy requires a semi-classical treatment of fluctuations around a curved background.  This approach share similarities with }
Note that there is another interesting analogy between different time scales: the Page time for black holes plays an analogous role to the quantum break time for de Sitter \cite{Zell}, and equivalently to the time scale where the back-reaction of cosmological perturbations on the cosmological background becomes important (see \cite{Abramo} for original work and \cite{RHBrev} for a review). 

The presence of the scrambling time indicates that an analysis beyond standard effective field theory of fluctuations is required in order to solve the black hole information problem. In a similar way,  our work lends support to the lesson from the TCC that a long lasting phase of accelerated expansion in cosmology can only be reliably analyzed if one goes beyond standard EFT.  For attempts to construct models of inflation going beyond an EFT treatment see e.g. \cite{Zell, Keshav}.

\section*{Acknowledgements}

R.B. is grateful for hospitality by the Institute for
Theoretical Physics and the Institute for Particle Physics and
Astrophysics of the ETH Zurich during the period when some of the work on
this project was carried out.  S.L. is supported in part by FRQNT.  The research at McGill is supported in part by funds from NSERC and from the Canada Research Chair program.

\appendix

%\section*{Appendix}

\end{document}